\documentclass[epj,twocolumn]{webofc}
\usepackage[varg]{txfonts}   
\usepackage{array} 
\usepackage{bm}
\usepackage{textcomp}
\usepackage{upgreek}

\wocname{\includegraphics[width=0.25cm,clip]{logo} ISVHECRI-2016}
\woctitle{ISVHECRI-2016}
\begin{document}
\title{Latest results of the Tunka Radio Extension}
%
%

\author{\firstname{D.}~\lastname{Kostunin}\inst{1}\fnsep\thanks{\email{dmitriy.kostunin@kit.edu}}
\and
\firstname{P.~A.}~\lastname{Bezyazeekov}\inst{2}
\and
\firstname{N.~M.}~\lastname{Budnev}\inst{2}
\and
\firstname{O.}~\lastname{Fedorov}\inst{2}
\and
\firstname{O.~A.}~\lastname{Gress}\inst{2}
\and
\firstname{A.}~\lastname{Haungs}\inst{1}
\and
\firstname{R.}~\lastname{Hiller}\inst{1}
\and
\firstname{T.}~\lastname{Huege}\inst{1}
\and
\firstname{Y.}~\lastname{Kazarina}\inst{2}
\and
\firstname{M.}~\lastname{Kleifges}\inst{3}
\and
\firstname{E.~E.}~\lastname{Korosteleva}\inst{4}
\and
\firstname{O.}~\lastname{Kr\"omer}\inst{3}
\and
\firstname{V.}~\lastname{Kungel}\inst{1}
\and
\firstname{L.~A.}~\lastname{Kuzmichev}\inst{4}
\and
\firstname{N.}~\lastname{Lubsandorzhiev}\inst{4}
\and
\firstname{T.}~\lastname{Marshalkina}\inst{2}
\and
\firstname{R.~R.}~\lastname{Mirgazov}\inst{2}
\and
\firstname{R.}~\lastname{Monkhoev}\inst{2}
\and
\firstname{E.~A.}~\lastname{Osipova}\inst{4}
\and
\firstname{A.}~\lastname{Pakhorukov}\inst{2}
\and
\firstname{L.}~\lastname{Pankov}\inst{2}
\and
\firstname{V.~V.}~\lastname{Prosin}\inst{4}
\and
\firstname{G.~I.}~\lastname{Rubtsov}\inst{5}
\and
\firstname{F.~G.}~\lastname{Schr\"oder}\inst{1}
\and
\firstname{R.}~\lastname{Wischnewski}\inst{6}
\and
\firstname{A.}~\lastname{Zagorodnikov}\inst{2}
~(Tunka-Rex Collaboration) 
}

\institute{
Institut f\"ur Kernphysik, Karlsruhe Institute of Technology (KIT), Karlsruhe, Germany  
\and
Institute of Applied Physics, Irkutsk State University (ISU), Irkutsk, Russia  
\and
Institut f\"ur Prozessdatenverarbeitung und Elektronik, Karlsruhe Institute of Technology (KIT), Germany
\and
Skobeltsyn Institute of Nuclear Physics, Lomonosov University (MSU), Moscow, Russia
\and
Institute for Nuclear Research of the Russian Academy of Sciences, Moscow, Russia  
\and
Deutsches Elektronen-Synchrotron (DESY), Zeuthen, Germany
}

\abstract{%
The Tunka Radio Extension (Tunka-Rex) is an antenna array consisting of 63 antennas at the location of the TAIGA facility (Tunka Advanced Instrument for cosmic ray physics and Gamma Astronomy) in Eastern Siberia, nearby Lake Baikal. 
Tunka-Rex is triggered by the air-Cherenkov array Tunka-133 during clear and moonless winter nights and by the scintillator array Tunka-Grande during the remaining time. 
Tunka-Rex measures the radio emission from the same air-showers as Tunka-133 and Tunka-Grande, but with a higher threshold of about 100 PeV. 
During the first stages of its operation, Tunka-Rex has proven, that sparse radio arrays can measure air-showers with an energy resolution of better than 15\% and the depth of the shower maximum with a resolution of better than 40 g/cm\textsuperscript{2}. 
To improve and interpret our measurements as well as to study systematic uncertainties due to interaction models, we perform radio simulations with CORSIKA and CoREAS. 
In this overview we present the setup of Tunka-Rex, discuss the achieved results and the prospects of mass-composition studies with radio arrays.
}
\maketitle
\section{Introduction}
The study of cosmic rays of ultra-high energies sheds light on the most powerful processes in the Universe.
The fine structures of the primary cosmic ray spectrum and the mass composition yield information on the type of cosmic accelerators and their location.
For example, at energies of about EeV, a transition from galactic to extragalactic cosmic ray sources is expected~\cite{Aloisio2012129}.
To distinguish between galactic and extragalactic sources, the precise determination of fluxes of different primary nuclei is required.
Modern optical detectors, namely, non-imaging air-Cherenkov arrays and fluorescence telescopes reach energy resolutions
of about 10\% and a resolution for the depth of the shower maximum ($X_\mathrm{max}$) of about 20~g/cm\textsuperscript{2}.
However, the duty cycle of such detectors is less than 15\%~\cite{Prosin:2016jev,Porcelli:2015jli}.

Digital radio arrays, as a novel technique, which allows for measurements of air-showers produced by primary cosmic rays with energies above 100~PeV.
A broad description of radio emission from air-showers, the technique of its detection and of historical and modern experiments is given in Ref.~\cite{SchroederReview2016}.

Modern detectors, such as LOFAR~\cite{Buitink:2016nkf}, AERA~\cite{Aab:2015vta} and Tunka-Rex~\cite{Bezyazeekov:2015ica} have already proven the feasibility of this technique, and shown that radio detection has a resolution competitive to optical technique.
Triggered by an external particle array, a radio detector becomes a scalable, cost-effective extension, which provides precise measurements of ultra-high energy cosmic rays around-the-clock.
In the present paper we focus on the setup of Tunka-Rex, discuss the achieved results and prospects of mass-composition studies with radio arrays.

\section{The Tunka Radio Extension}
\begin{figure}[t]
\includegraphics[width=1.0\linewidth]{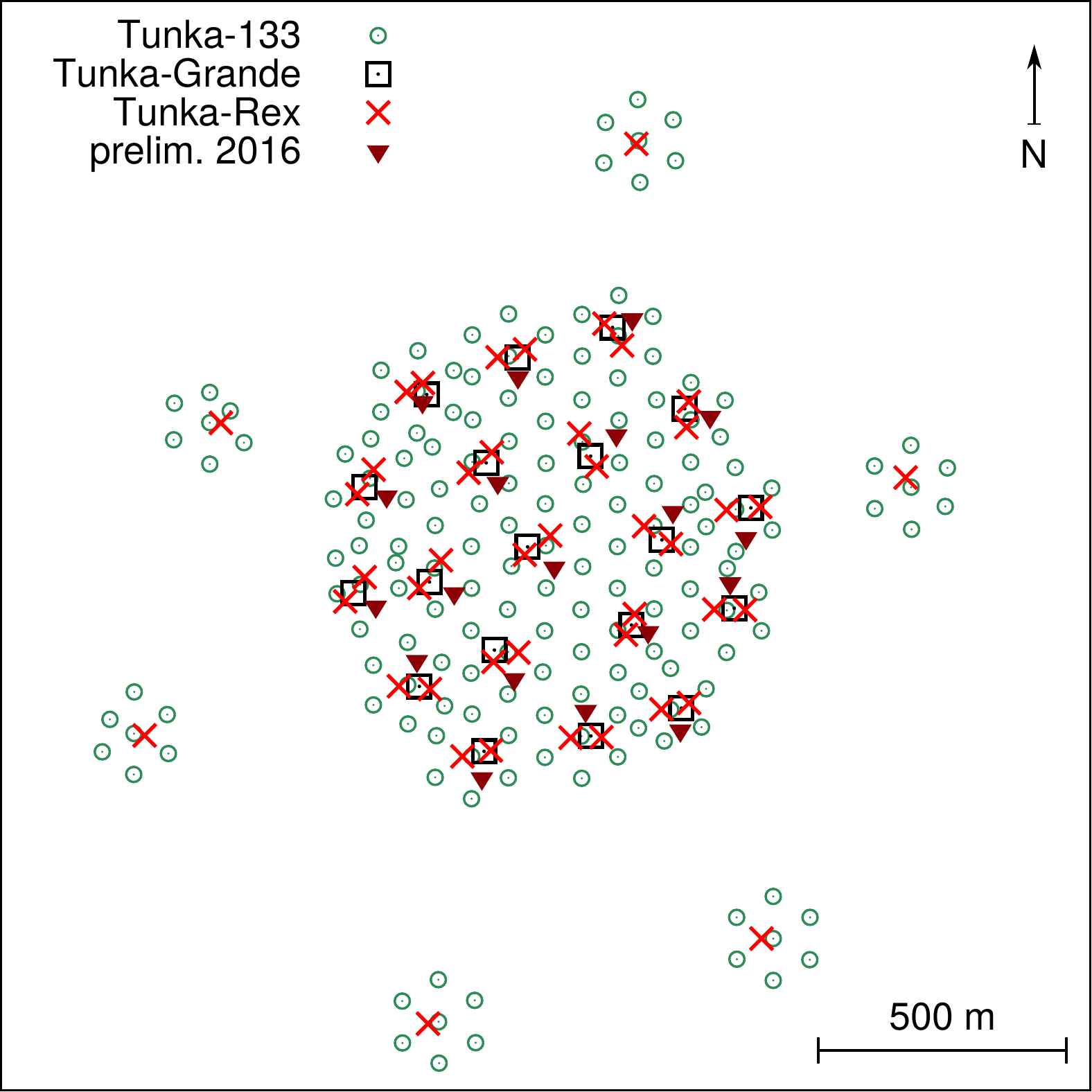}
\caption{Layout of cosmic-ray detectors of the TAIGA~\cite{Budnev:2016btu} facility. 
The core consists of 19 clusters, each of them is equipped with 3 Tunka-Rex antenna station, and 6 satellite clusters with one Tunka-Rex antenna station per cluster.
Triangles depict preliminary positions of the antennas deployed in 2016 (precise position measurements will be performed soon).
}
\label{fig:layout}
\end{figure}

The Tunka Radio Extension (Tunka-Rex) array has been commissioned in 2012 and originally consisted of 18 antennas distributed over an area of 1 km\textsuperscript{2}~\cite{TunkaRex_NIM_2015}.
The detector layout is mostly determined by the Tunka-133~\cite{Prosin:2016jev} clusters, the original air-Cherenkov array of the TAIGA facility~\cite{Budnev:2016btu} located nearby southern tip of Lake Baikal in Siberia.
At the moment the cosmic-ray detectors of TAIGA consists of Tunka-133, Tunka-Grande~\cite{Budnev:2015cha} and Tunka-Rex.
Tunka-Rex now contains 63 antenna stations including six satellite stations extending the area to 3~km\textsuperscript{2}.
The common layout of the three experiments is shown in Fig.~\ref{fig:layout}.

\begin{figure}[t]
\includegraphics[width=1.0\linewidth]{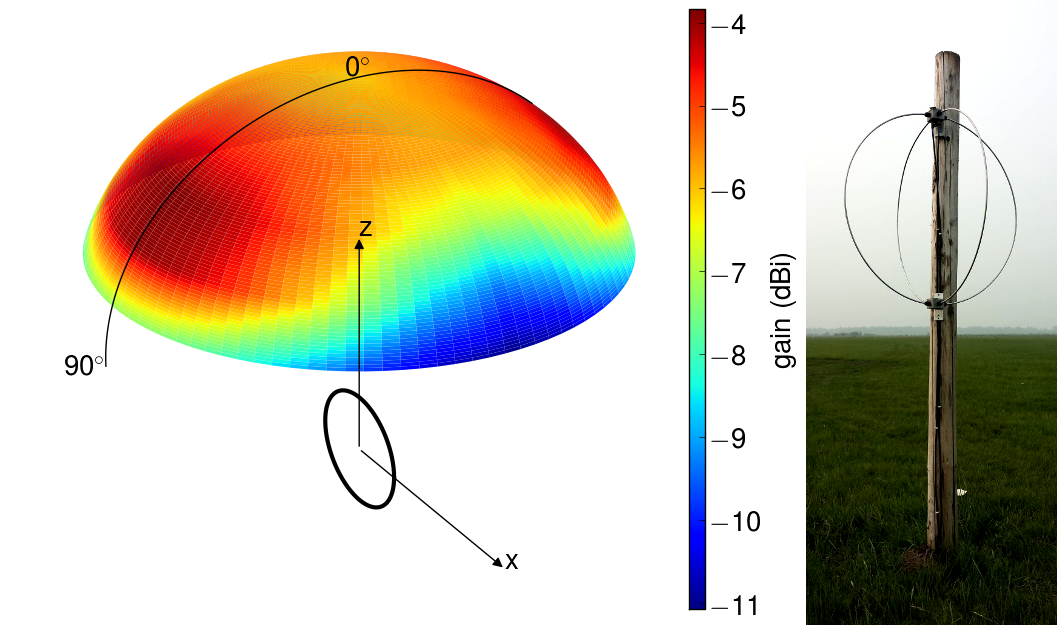}
\caption{
\textit{Left: }Gain pattern of the SALLA antennas of Tunka-Rex.
\textit{Right: }A Tunka-Rex antenna station. Two perpendicular SALLAs are mounted on a wooden pole at the height of about 3 m (upper end of SALLA).
}
\label{fig:antenna}
\end{figure}

Each Tunka-Rex antenna station consists of two perpendicular short aperiodic loaded loop antennas~(SALLA)~\cite{KroemerSALLAIcrc2009,Abreu:2012pi} rotated by an angle of $\pm$45$^\circ$ with respect to magnetic North.
Each antenna contains a low-noise amplifier (LNA) and a load suppressing the downward direction, 
which makes the antenna upward-looking and decreases the uncertainty due to ground conditions to the level of only 3\%.
A Tunka-Rex antenna station and the SALLA gain pattern are shown in Fig.~\ref{fig:antenna}.
Before digitalization, the signals are analogically processed with a filter-amplifier with an effective band of 30-76~MHz.
Each Tunka-Rex antenna station is connected either to the Tunka-133 or the Tunka-Grande local data acquisition and shares the same ADC boards.

All clusters are synchronized with the central DAQ via optical fibers.
We have checked the stability of the synchronization with a beacon-based method~\cite{Schroeder2010277,PierreAuger:2016zxi}.
The relative timing is stable to better than a nanosecond during a single run, however, after reset we obtain jumps of about 5~ns.
Taking background into account, the resulting timing uncertainty is about 7~ns.
Depending on the trigger mode, the entire cluster is triggered by air-Cherenkov (clear winter moonless nights) or particle detectors (the rest of the time),
and traces from the Tunka-133 PMTs (when operating), scintillators and antennas are recorded simultaneously in traces of 1024 samples with 5~ns sampling rate and a bitdepth of 12 bits.
As a result, TAIGA features duplex (particles + radio) and triplex (particles + radio + air-Cherenkov) measurements of cosmic rays with energies of $10^{16}$--$10^{18}$~eV.

\subsection{Antenna calibration}
The calibration of the Tunka-Rex antenna was performed in the following way: the directivity of the SALLA antenna was simulated with the NEC2 code~\cite{nec2}, 
and then normalized to an amplitude calibration made with the commercial reference source VSQ1000+DPA4000 by Schaffner Electrotest GmbH (now Teseq).
The same method was also used to calibrate the LOPES antennas~\cite{LOPES:2015eya}, as well as the LOFAR ones~\cite{Nelles:2015gca}, what makes these three experiments having consistent absolute calibration scale.
The hardware response and temperature dependence of the LNA and filter-amplifier were measured under laboratory conditions, and the calibration of the ADC was done on the board already deployed at the local DAQ of the clusters.
As result, the overall uncertainty on the absolute amplitude reconstruction is 22\%, with a dominating contribution of 16\% from the calibration scale uncertainty of the reference source, and a number of smaller contributions given by environmental conditions, antenna production and deployment.

\section{Event reconstruction}
\begin{figure*}[t]
\begin{center}
\includegraphics[height=0.26\linewidth]{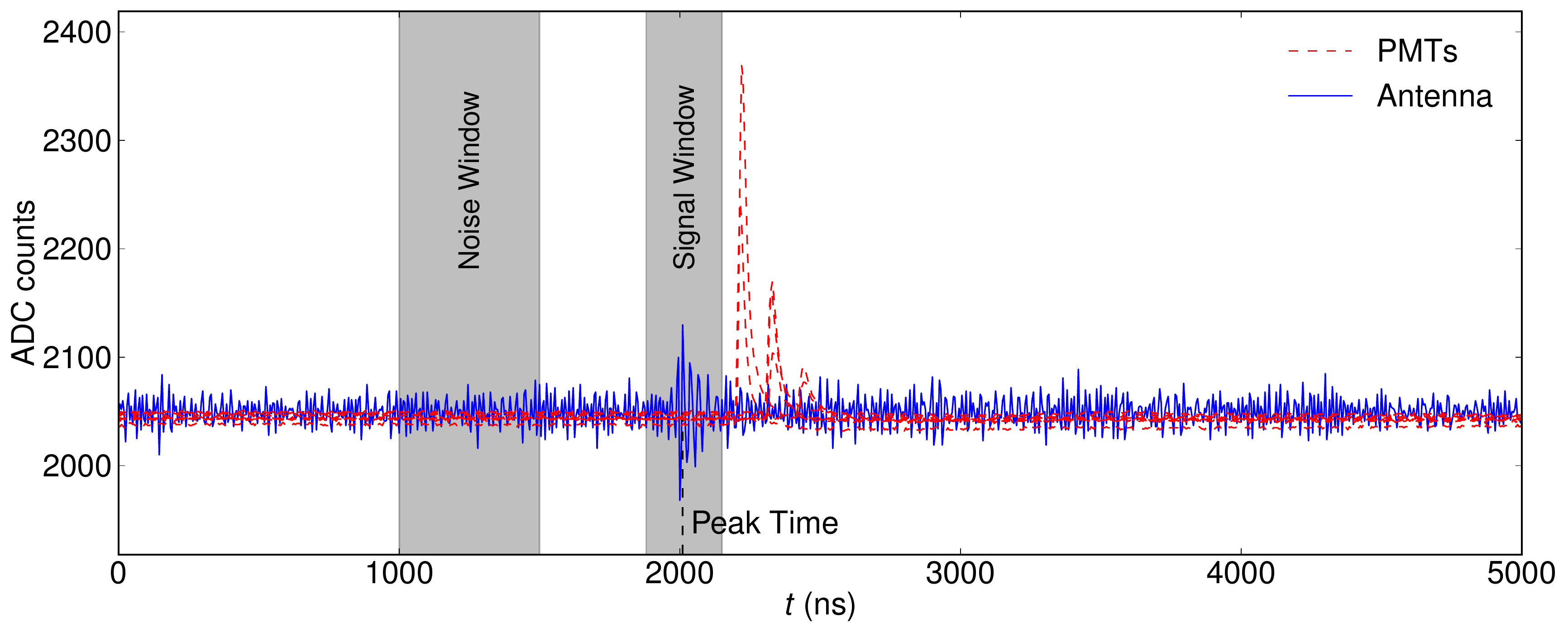}~~~~\includegraphics[height=0.26\linewidth]{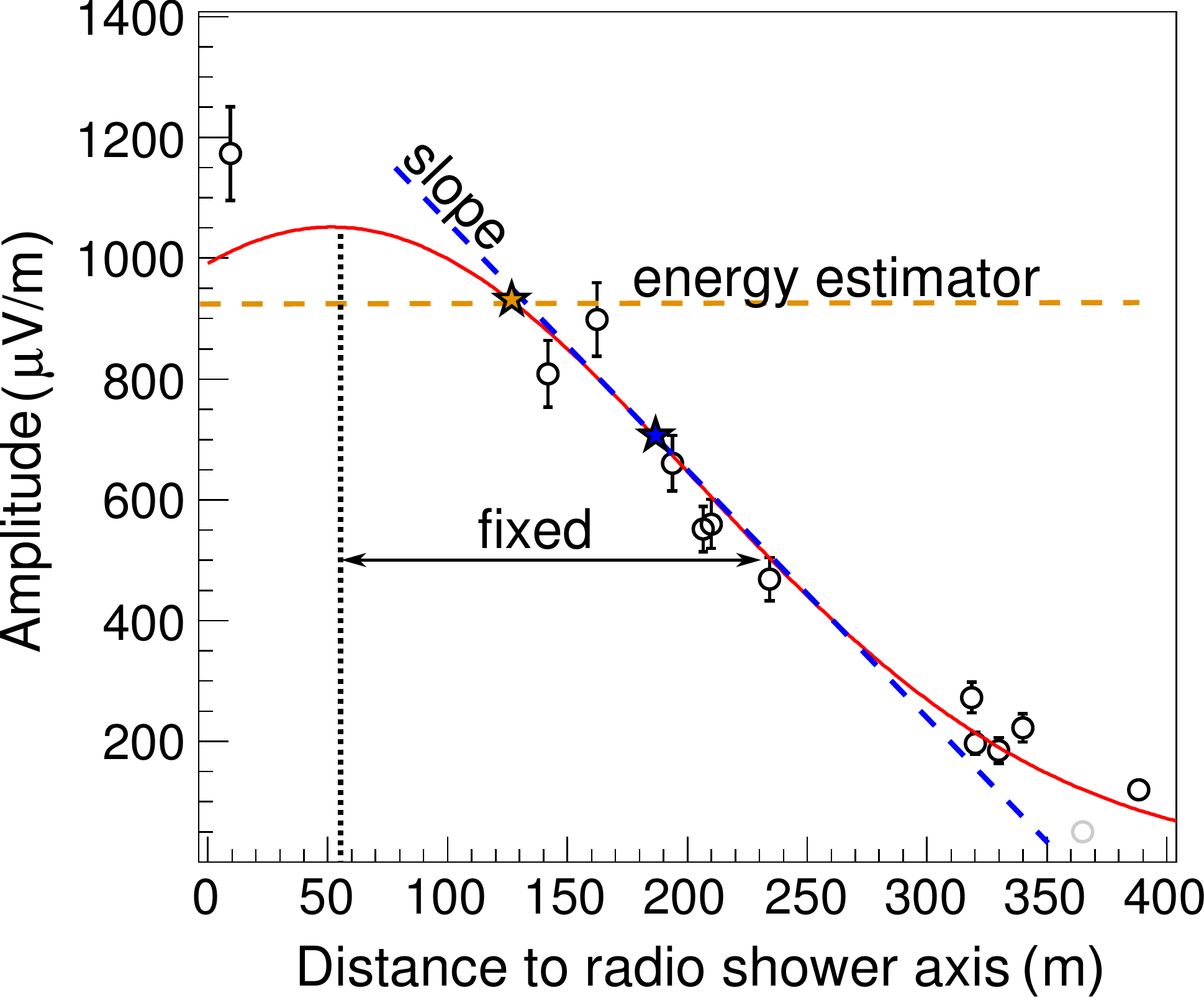}
\caption{\textit{Left: }Example traces recorded by the Tunka-133 local DAQ. 
The radio signal is recorded earlier than the PMT pulses, mostly due to longer cables of Tunka-133.
\textit{Right: }Sketch describing the reconstruction of the primary energy and distance to the shower maximum. 
The lateral distribution is fitted by a Gaussian-shape function with fixed width. 
The primary energy is proportional to the amplitude at~120~m distance, and the distance to the shower maximum is calculated from the slope at~180~m from the shower axis.
}
\label{fig:trace-recoldf}
\end{center}
\end{figure*}

Since all clusters operate independently, single-cluster events are merged into shower events during offline analysis.
At the first step, only events containing at least three antenna stations of signal-to-noise ratio~(SNR) in power ${S^2/N^2 > 10}$ are selected.
The amplitude of the signal $S$ is defined as the maximum of amplitude of the Hilbert envelope of the vectorial sum of the two measured polarizations inside of the signal window.
The position of the signal window is constant and defined by the hardware delay between the radio signal arrival time and the particle or air-Cherenkov trigger,
while the width of the signal window is defined by timing uncertainties and different shower geometries.
The noise level~$N$ is defined as RMS of the amplitude in a noise window.
An example of radio and air-Cherenkov traces recorded at the same cluster are given in Fig.~\ref{fig:trace-recoldf}.
Since Tunka-Rex is operating close to the threshold, the contribution of the background cannot be neglected.
This contribution is taken into account for the estimation of timing and amplitude uncertainties, 
and the measured amplitudes is corrected for a systematic background bias using a function depending on SNR.

At the next step, the arrival direction is reconstructed with a plane wave front model, and is compared to the one measured by the master detector (Tunka-133/Tunka-Grande).
Since the angular resolution of Tunka-Rex is about 1--2$^\circ$, all events with direction deviating from the master reconstruction by more than 5$^\circ$ are rejected and excluded from analysis.

After the first quality cuts, the position of the shower core is reconstructed.
In the triplex mode, the shower core and axis is taken from the Tunka-133 reconstruction, since it has much larger density than the other cosmic-ray detectors of TAIGA, and the resolution is better than 5~m.
In the duplex mode we plan to combine the reconstructions from Tunka-Rex and Tunka-Grande, since both of them are expected to feature a resolution of about 20-30~m due to their spacing, 
and we expect that the combined reconstruction will improve this value.

After the reconstruction of the shower core, the amplitudes from the detector surface are projected to the shower axis forming the lateral distribution.
Any antenna station passed SNR cut appears further on the lateral distribution than two antenna stations without the signal is considered as outlier and rejected as false positive.

To reconstruct the primary energy $E_\mathrm{pr}$ and the depth of the shower maximum $X_\mathrm{max}$, the lateral distribution is corrected to remove the dependence on a geomagnetic and azimuth angles, which is introduced by the interference of the geomagnetic and charge excess effects~\cite{Kostunin:2015taa}.
The resulting distribution is fitted with a lateral distribution function (LDF), containing two free parameters.
These two parameters, namely normalization and slope, are used for the reconstruction of $E_\mathrm{pr}$ and $X_\mathrm{max}$, respectively.
A sketch describing the idea of the LDF approach is given in Fig.~\ref{fig:trace-recoldf}.
Since $X_\mathrm{max}$ is very sensitive to the shape of the LDF, we apply additional quality cuts to select high-quality events: 
the event must contain at least one antenna further than 200~m from the shower axis (to increase the sensitivity to the LDF slope), and the resulting fit uncertainty of $X_\mathrm{max}$ must be less than~50~g/cm\textsuperscript{2}.

It is worth noting, that radio technique is sensitive only to electromagnetic component of air-showers, 
which means that method have additional uncertainties due to unknown primary particles.
These uncertainties is discussed in section~\ref{sec:syst_unc}.

\begin{figure}[t]
\includegraphics[width=1.0\linewidth]{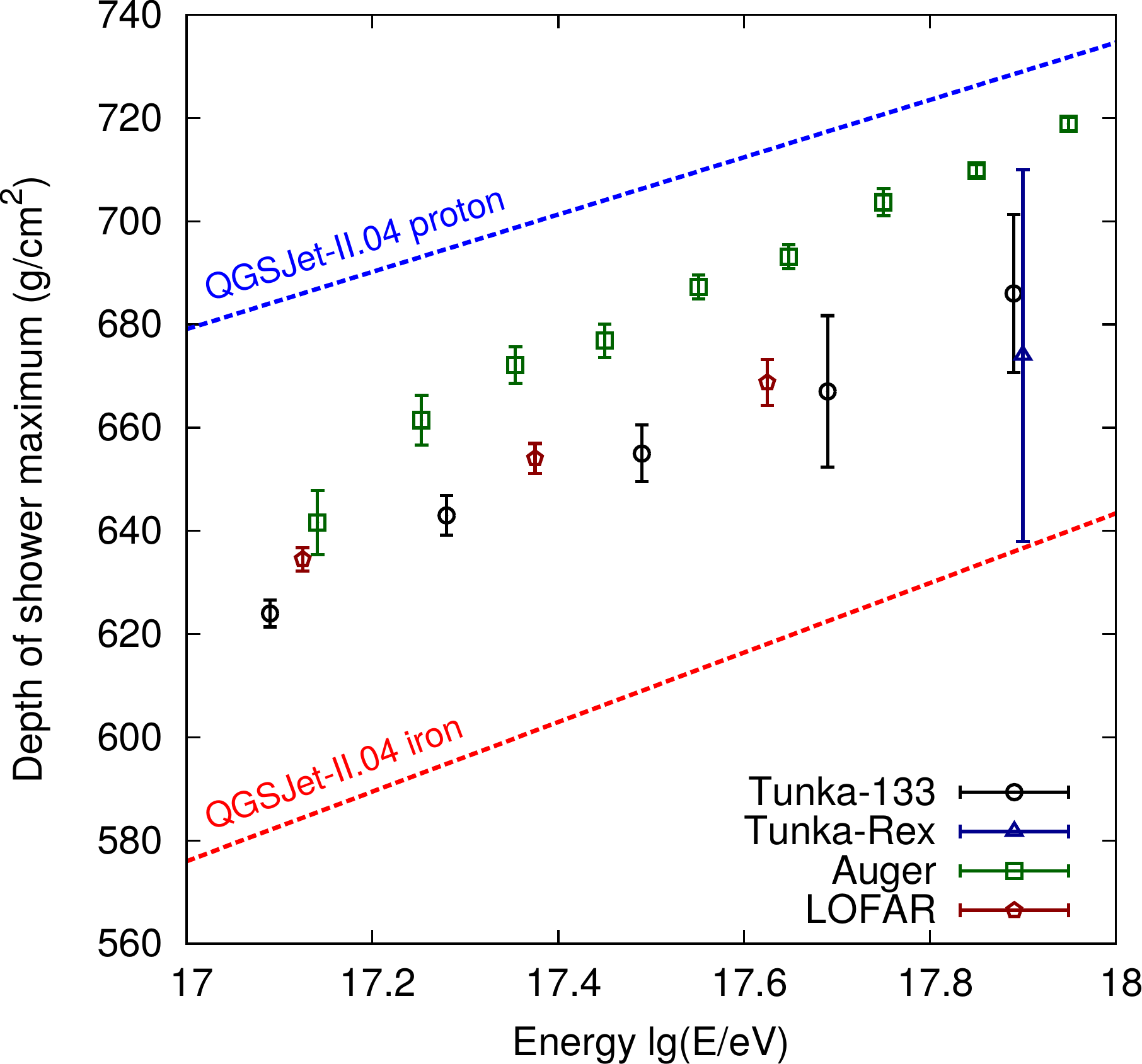}
\caption{
Mean depth of the shower maximum versus the primary energy reconstructed by modern cosmic-ray experiments: Tunka-133~\cite{Prosin:2016jev}, Auger~\cite{AugerHEATXmaxICRC2015}, LOFAR~\cite{Buitink:2016nkf} and Tunka-Rex.
Errorbars indicate statistic uncertainties only, and all measurements agree within additional systematic uncertainties.
}
\label{fig:xmax}
\end{figure}

Up to now, we have finished the reconstruction of measurements during 2012-2014, when Tunka-Rex was operating jointly with Tunka-133.
To compare the reconstructions of the detectors, 
we selected events with core positions inside the dense part of detector, i.e. within 500~m radius around the center.
To avoid implicit tuning in the cross-check of Tunka-Rex and Tunka-133, the half of Tunka-133 reconstruction of $E_\mathrm{pr}$ and $X_\mathrm{max}$ was blinded, and opened only after the final reconstruction of Tunka-Rex.
The comparing set includes 148 events with reconstructed energy and 42 events with reconstructed $X_\mathrm{max}$.
The obtained resolution of Tunka-Rex is 15\% for energy and 40~g/cm\textsuperscript{2} for $X_\mathrm{max}$, while no significant absolute shift between the reconstructions of Tunka-Rex and Tunka-133 was observed.
Since all of the high-energy events have passed the quality cuts, we can use them for a mass-composition study. 
The mean $X_\mathrm{max}$ value obtained for ${\lg(E_\mathrm{pr}/\mbox{eV}) = 17.9\pm0.1}$ is given in Fig.~\ref{fig:xmax} based on 8 events. 

It is worth noting, that these results were obtained with the Tunka-Rex configuration consisting of one antenna per cluster.
Meanwhile, starting from 2016 the array features three antennas per cluster.

To better understand systematic uncertainties, atmospheric effects will have to be taken into account~\cite{Abraham:2009bc}.
For the effective frequencies of Tunka-Rex these uncertainties can be in the order of 2\% and 5~g/cm\textsuperscript{2} for the energy~\cite{Glaser:2016qso} and shower maximum~\cite{Buitink:2016nkf} reconstructions, respectively.
In future, we plan to include Global Data Assimilation System (GDAS) data to our analysis to decrease these uncertainties~\cite{Abreu:2012zg}.

\subsection{Energy reconstruction with a single antenna station}
Besides the main goals of energy and shower maximum reconstruction we have shown that a single antenna station can provide useful information when the shower core and axis are known~\cite{HillerArena2016}.
Assuming a mean value for the steepness of the radio LDF of about ${\eta_0^{-1} \approx 120}$~m,
and a threshold of ${S_\mathrm{th} \approx 90}$~{\textmu}V/m,
using reconstructed shower geometry from Tunka-133, we obtained a reliable energy reconstruction using single-antenna events.
Particularly, the number of events was increased by more than three times, while the energy resolution has slightly decreased to 20\%.

This result indicates the feasibility of equipping surface particle detectors with simple radio extensions, which allow for the determination of the electromagnetic energy deposit of high-energy air-showers above $10^{17}$~eV.

\section{The radio amplitude as measure for the absolute energy scale}
The independent energy reconstruction by Tunka-Rex is based on an absolute amplitude calibration of the antennas and on normalization parameters obtained with CoREAS.
As shown above, that the absolute energy scales of Tunka-Rex and Tunka-133 experiments are in very good agreement.
Since the radio emission from air-showers is well understood and its generation and propagation depend weakly on the atmospheric condition, it can be used as universal tool to compare or cross-check the energy scale of experiments located in different places and exploiting different techniques.
To test this statement, we have selected KASCADE-Grande~\cite{Apel2012183} with its radio extension LOPES~\cite{Apel:2014ika} and compared it with Tunka-133 and Tunka-Rex measurement, respectively~\cite{Apel:2016gws}.
Since Tunka-Rex and LOPES were calibrated with the same reference source,
most systematic uncertainties of the amplitude calibration cancel out in the comparison.

We realize this comparison with two different approaches.
The first one is to compare the ratio $\kappa$ of measured radio amplitudes and the energy reconstructed by the master experiment.
Then the relative shift between the masters is defined as ${f_\mathrm{amp} = \kappa_{\mathrm{Tunka-Rex}}/\kappa_{\mathrm{LOPES}}}$.
This method relies only on direct radio measurements, 
and the reconstruction procedure are chosen as similar as possible (i.e. the same bandwidth and the same LDF treatment), 
moreover, the reconstruction is corrected for the different observation depths and magnetic fields of the two locations.

The second method is implemented via CoREAS simulations: two simulations -- one with proton and one with iron primary were produced for each event, and then the measured and simulated radio amplitudes were compared to each other.
Then, the mean ratios $F^\mathrm{p}$ ($F^\mathrm{Fe}$) between simulated and measured amplitudes are determined, and the scale shift between the master experiments is defined as ${f_\mathrm{sim} = F_{\mathrm{Tunka-Rex}}/F_{\mathrm{LOPES}}}$.
The main uncertainty of these method is given by the hadronic model used in the simulation and uncertainties due to angular dependence of antenna gain.

\begin{figure}[t]
\includegraphics[width=1.0\linewidth]{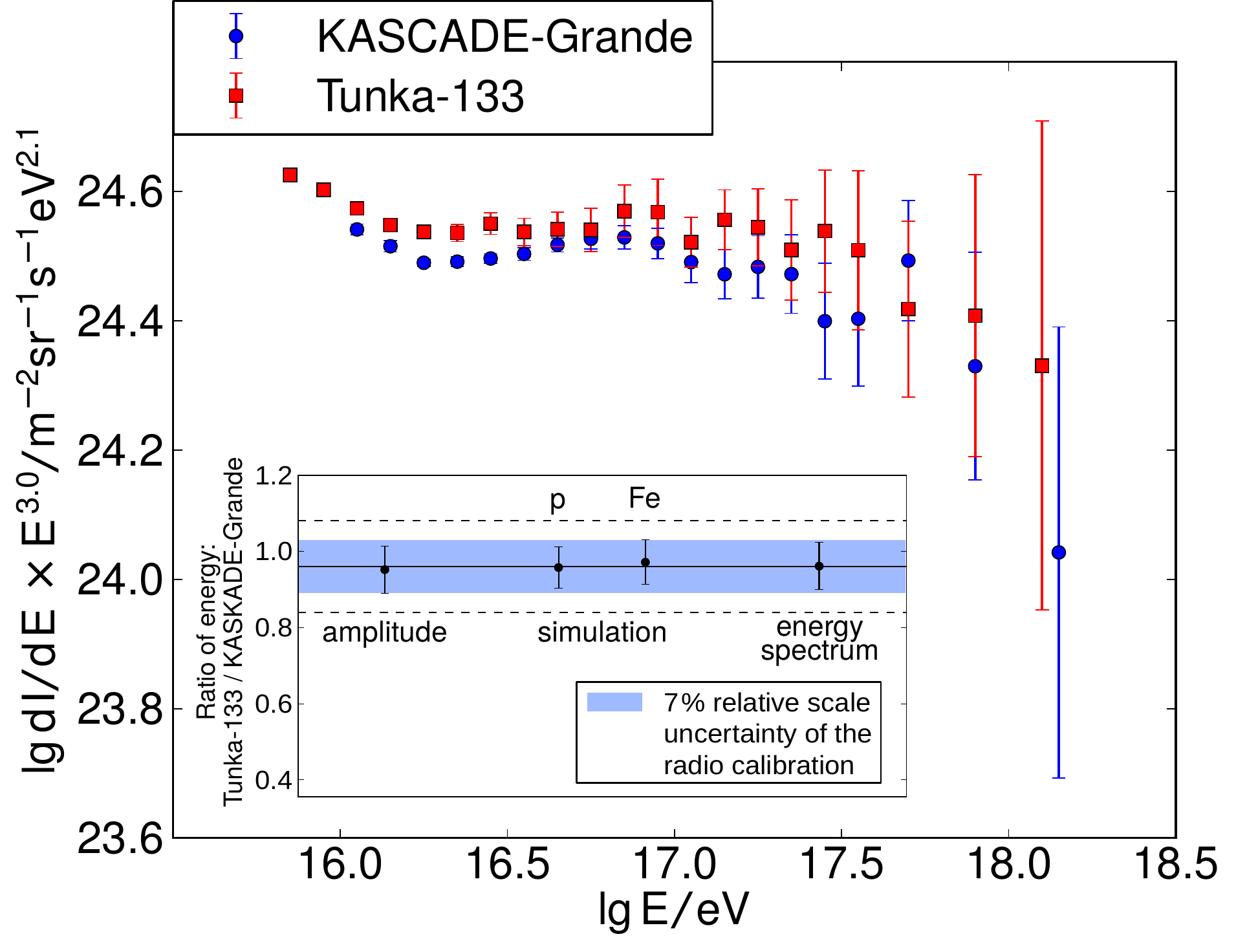}
\caption{
Energy spectra of cosmic rays from KASCADE-Grande~\cite{Apel2012183} and Tunka-133~\cite{Prosin:2016jev}: normalized flux per energy;
and the results from the comparison of the energy scales between the experiments Tunka-Rex and LOPES (small box) in the energy range of $10^{17}$ to $10^{18}$\,eV.
With a systematic increase of KASCADE-Grande energies by $4\,\%$ (or a corresponding decrease of Tunka-133 energies) the average flux per energy of both experiments can be brought to agreement in this energy range~\cite{Apel:2016gws}. 
}
\label{fig:spec_comparison}
\end{figure}

The result of both methods is that the energy scales of Tunka-133 and KASCADE-Grande are consistent to about 10\% limited by systematic uncertainties of the LOPES experiment, and the mean KASCADE-Grande energy scale is lower than Tunka-133 by about 5\%.
A similar shift is obtained by a straight-forward fit of the energy spectra of Tunka-133 and KASCADE-Grande.
The spectra and results of the scale comparison are given in Fig.~\ref{fig:spec_comparison}.

This result can be applied to study finer features of the energy spectrum with higher accuracy, e.g. it allows to define the positions of knee-like structures with lower uncertainty.

\section{Ongoing activity}
In this section we discuss theoretical work performing in the frame of Tunka-Rex experiment with a purpose of improving the reconstruction of air-shower events.

\subsection{Lateral distribution function}
Hereafter we perform calculations in the frame of the geomagnetic coordinate system based on the shower axis vector $\bm{\hat{\mathrm{V}}}$ and the vector of the Earth's magnetic field $\bm{\hat{\mathrm{B}}}$ (a hat over a vector denotes a unit vector: $\bm{\hat{\mathrm{B}}} = \bm{\mathrm{B}} / |\bm{\mathrm{B}}|$):
\begin{eqnarray}
&& \bm{\hat{\mathrm{e}}}_x = \bm{\hat{\mathrm{V}}} \times \bm{\hat{\mathrm{B}}}\,, \\
&& \bm{\hat{\mathrm{e}}}_y = \bm{\hat{\mathrm{V}}} \times (\bm{\hat{\mathrm{V}}} \times \bm{\hat{\mathrm{B}}})\,, \\
&& \bm{\hat{\mathrm{e}}}_z = \bm{\hat{\mathrm{V}}}\,.
\end{eqnarray}
This way, the shower front is laying in the plane $(\bm{\hat{\mathrm{e}}}_x, \bm{\hat{\mathrm{e}}}_y)$.
Let us also define useful angles: the geomagnetic angle between shower axis and magnetic field $\alpha_{\mathrm g} = \angle(\bm{\mathrm{V}},\bm{\mathrm{B}})$ and the geomagnetic azimuth $\phi_{\mathrm g} = \angle(\bm{\hat{\mathrm{e}}}_x,\bm{\mathrm{r}})$,
where $\bm{\mathrm{r}}$ is the coordinate of an antenna station.

The main parameterization used by Tunka-Rex is the parameterization describing the distribution of radio amplitudes with respect to the shower axis, i.e. the LDF:
\begin{eqnarray}
&&\mathcal{E}(r,\phi_\mathrm{g}) = \mathsf{\hat K}^{-1}(\phi_\mathrm{g})\mathcal{E}_2(r), \\
&&\mathsf{\hat K}(\phi_\mathrm{g}) = \left(\varepsilon^2 + 2\varepsilon\cos\phi_{\mathrm g}\sin\alpha_{\mathrm g} + \sin^2\alpha_{\mathrm g}\right)^{-\frac{1}{2}}, \\
&&\mathcal{E}_2(r) = \mathcal{E}_0\exp\left(a_1(r-r_0) + a_2(r-r_0)^2\right),
\end{eqnarray}
where $\mathcal{E}(r,\phi_\mathrm{g})$ is the amplitude at the antenna station with coordinates $(r,\phi_\mathrm{g})$.
This amplitude is described with two significant terms: first, the azimuthal asymmetry $\varepsilon$ is corrected by the term $\mathsf{\hat K}(\phi_\mathrm{g})$; 
second, the lateral distribution $\mathcal{E}_2(r)$, 
which is a Gaussian-like function with the normalization $\mathcal{E}_0$, with a width proportional to $a_2$ and the slope proportional to $a_1$.
The maximum of $\mathcal{E}_2(r)$ reflects Cherenkov-like features of the radio emission.
The parameter $r_0$ is arbitrarily chosen in a way to obtain maximum correlation of $\mathcal{E}_0$ and $a_1$ with the primary energy and the distance to the shower maximum, respectively.

\subsection{Estimation of core position}
After the upgrade of Tunka-Rex, we are now implementing an independent reconstruction of the position of shower core with radio standalone.
The position of the core $(x_0,y_0)$ is added to LDF parameters ${r = r(x_0,y_0)}$ and $\phi_\mathrm{g} = \phi_\mathrm{g}(x_0,y_0)$.
Since Tunka-Rex operates near the threshold, currently the core position is reconstructed in three steps:
\begin{enumerate}
\item The initial core position is estimated a center of mass of the radio amplitudes during the arrival-direction reconstruction (requires at least 3 antennas);
\item Parameters $a_1$ and $a_2$ are fixed to default values, and the LDF is fitted with three free parameters: $\mathcal{E}_0$, $x_0$ and $y_0$ (requires at least 4 antennas);
\item The LDF is fitted again including $a_1$ and $a_2$, and the limits for $\mathcal{E}_0$, $x_0$ and $y_0$ are defined from the previous stage (requires at least 6 antennas).
\end{enumerate}
With this procedure we expect to obtain resolutions of about 20-30~m for the dense part of the detector (inner circle of 500~m around the center of the array).
To improve these numbers we plan to apply stricter quality cuts on the signal reconstruction for this particular procedure.

\subsection{Limitation of the one-dimensional approach}
For the time being, all methods for the $X_\mathrm{max}$ reconstruction are based on the simple relation between single parameters of the LDF and $X_\mathrm{max}$ or the distance to $X_\mathrm{max}$.
For example, Tunka-Rex, LOFAR~\cite{Nelles:2014xaa} and AERA use Gaussian-like parameterizations and exploit the correlation of $X_\mathrm{max}$ with the slope and width of LDF, respectively.

In this section we describe the relation between the position of the depth of the shower maximum and the slope of the LDF,
give a more strict consideration of it,
and discuss its possible hidden features.

We use the following assumptions: the distribution of the electrons behaves as Gaisser-Hillas function and the density of the Earth's atmosphere falls exponentially with increasing altitude, namely we use the CORSIKA parameterization of the standard atmosphere~\cite{HeckKnappCapdevielle1998}.
The simple form of the amplitude of the radio signal $\mathcal{E}_\nu(r)$ with frequency $\nu$ at distance $r$ from the shower axis is~\cite{Allan1971}:
\begin{equation}
\label{eq1}
\mathcal{E}_\nu(r) = \kappa \int\limits^{h^\nu_2(r,n_\mathrm{r})}_{h^\nu_1(r,n_\mathrm{r})} \frac{N(h)}{h} \mathrm{d}h\,,
\end{equation}
where $\kappa$ is the normalization coefficient (the dependence on geomagnetic angle $\alpha_\mathrm{g}$ has already been taken into account),
$N(h)$ is the number of electrons at the altitude $h$, and $h^\nu_{1,2}(r,n_\mathrm{r})$ are the integration limits depending on the distance to shower axis $r$ and refractive index $n_\mathrm{r}$.

Assuming that the refractive index $n_\mathrm{r}(h)$ is proportional to the density of the atmosphere, one can recalculate it from the atmospheric parameterization and use it as an input to estimate the integration limits $h^\nu_{1,2}(r,n_\mathrm{r})$ for \textit{vertical} air-showers:
\begin{equation}
h_{1,2}^\nu(r,n_\mathrm{r}) = \left(\frac{r}{r_{1,2}(\nu,n_\mathrm{r})}\right)^{\alpha_{1,2}(\nu,n_\mathrm{r})}\,,
\label{eq:h12par}
\end{equation}
This parameterization is obtained by numerical solution of the equation
\begin{equation}
\Delta t (r,h^\nu) = \Delta t (r,h_c^\nu) + \frac{1}{2\pi\nu}\,,
\end{equation}
where 
\begin{equation}
\Delta t (r,h^\nu) = \cos\left(\arctan\left(\frac{r}{h^\nu}\right)\right)\int\limits_0^{h^\nu}\frac{n_\mathrm{r}(h')}{c}\mathrm{d}h' - \frac{h^\nu}{c}\,,
\end{equation}
where $c$ is the velocity of light, and $h_c$ is defined as solution of equation
\begin{equation}
\frac{\partial}{\partial h_c} \Delta t (r,h_c) = 0\,.
\end{equation}
For inclined air-showers everything is scaled by $\cos\theta$ (where $\theta$ is the zenith angle) at first approximation, but here we do not consider these cases.

\begin{figure*}
\begin{center}
\includegraphics[height=0.43\linewidth]{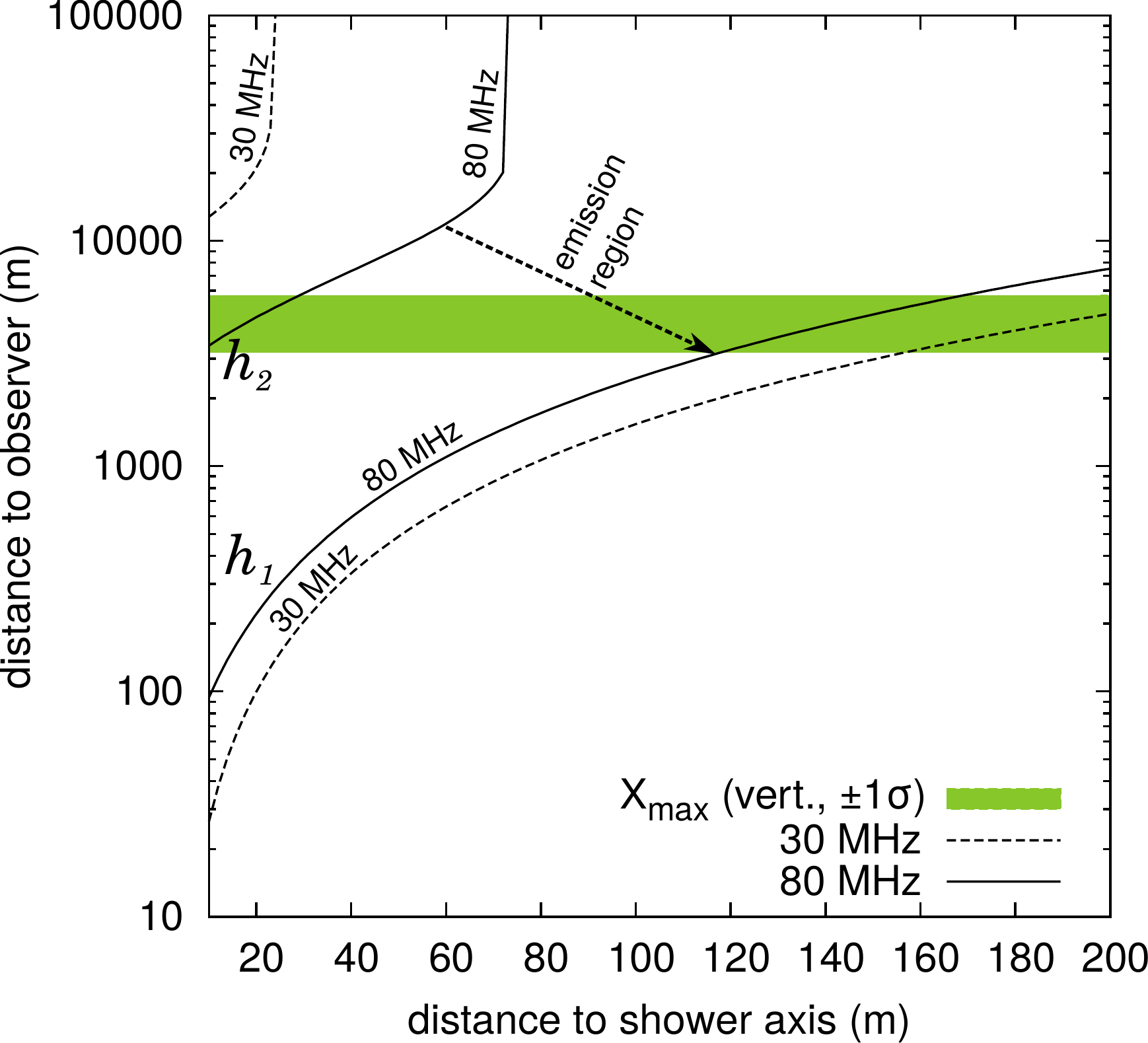}~~~~~\includegraphics[height=0.43\linewidth]{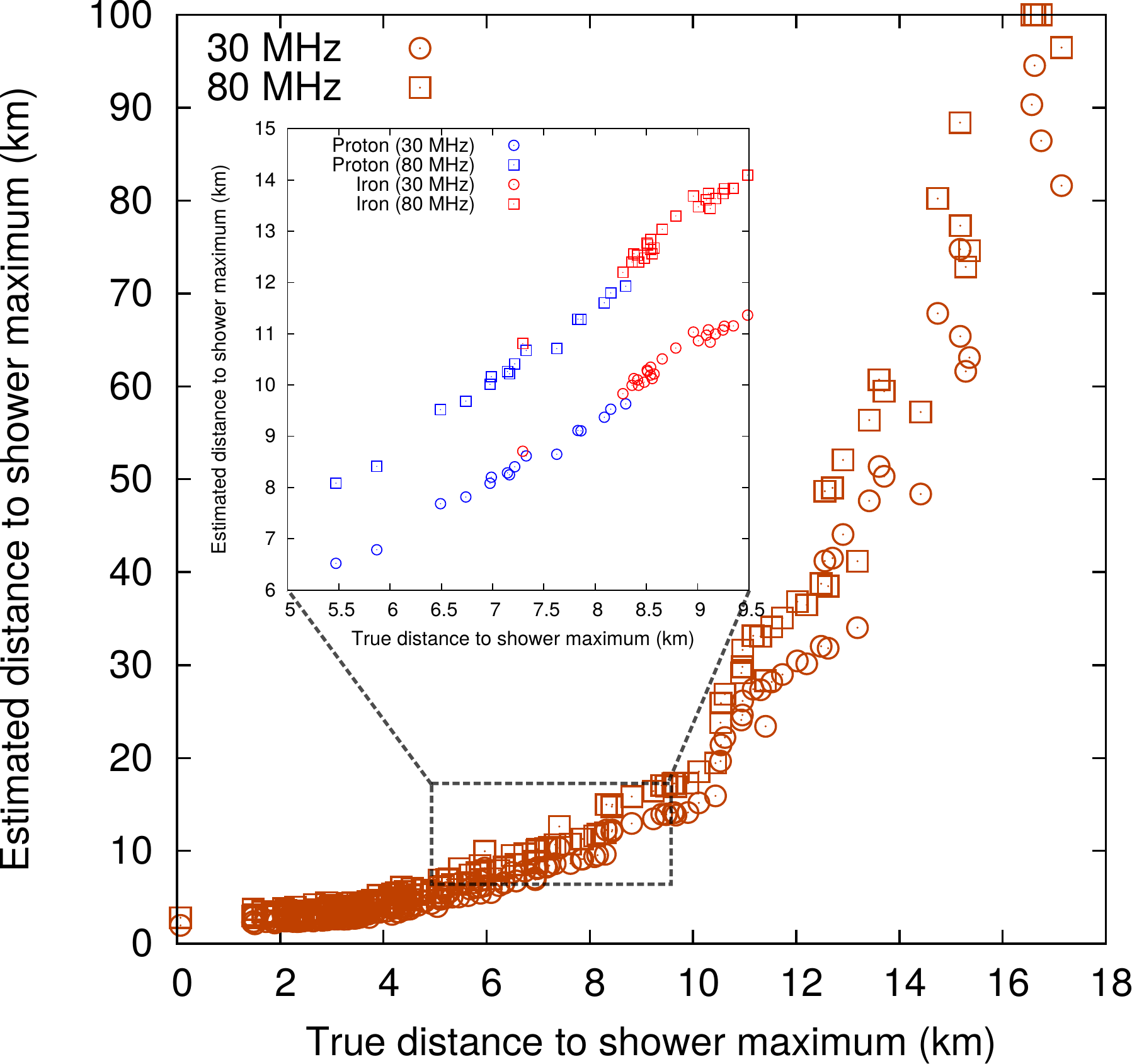}
\caption{\textit{Left: }Sketch describing the radio emission from air-showers.
The X axis indicates the position of the observing antenna, the Y axis indicates the altitude (the observer is placed on zero altitude),
the colored band indicates possible positions of the shower maximum for typical air-showers (one sigma spread around the mean due to shower-to-shower fluctuations for vertical air-showers of energies $10^{17}$ -- $10^{18}$~eV), and the lines indicate the bounds of the emission region for different frequencies.
These lines are calculated using the atmosphere refractivity $n_\mathrm{r}$ as input.
One can see, that the emission region reduces with increasing frequency and the point of intersection with the shower maximum will define the Cherenkov ring observed for high frequencies.
For lower frequencies one can see the exponential fall-off of amplitudes after the intersection of $h_1$ with the shower maximum line.\hspace{\textwidth}
\textit{Right: }Correlation between the true geometrical distance to the shower maximum and the estimation made with the simplest slope method neglecting particle interactions and propagation constants.
We obtained no difference between slopes of proton and iron induced air-showers.
The dependence is not linear mostly due to the different density of the atmosphere at the different altitudes (i.e. the geometrical mean path length of electron is different); for real data analysis we use a more sophisticaed method tuning all reconstruction constants are tuned against CoREAS simulations to obtain a linear dependence~\cite{Kostunin:2015taa,Bezyazeekov:2015ica}.
} 
\label{fig:cr-xmax-geo}
\end{center}
\end{figure*}

The curves denoting the behavior of $h^\nu_{1,2}(r,n_\mathrm{r})$ are in Fig~\ref{fig:cr-xmax-geo}.
One can see that for lateral distances far from the Cherenkov bump (${r>100}$~m), the upper limit goes to infinity (becomes higher than the top of the atmosphere) and the lower limit goes to the position of the shower maximum and above.

In our case, the integration limits in~Eq.~(\ref{eq1}) are as follows: $h^\nu_2(r,n_\mathrm{r})\to\infty$, $h^\nu_1(r,n_\mathrm{r}) > h_\mathrm{max}$, i.e. we integrate over the upper tail of the electron distribution.
This way, the value of the integral~Eq.~(\ref{eq1}) has the following form:

\begin{equation}
\mathcal{E}_\nu(r > r_\mathrm{c}) \propto \exp \left(-\frac{(r/r_1)^{\alpha_1}}{h_\mathrm{max}}f_{\mathrm{int}}(h_\mathrm{max},...)\right)\,.
\end{equation}
As it was expected, we obtained an exponential-like behavior of the LDF, where the exponent is defined by the altitude of the shower maximum $h_\mathrm{max}$.
It is worth noting that the power of $r$ (defined as $\alpha_1 \approx 3/2$, which is obtained from the fit of~Eq.~(\ref{eq:h12par}) to curves shown in Fig.~\ref{fig:cr-xmax-geo}) is between 1 and 2, i.e. both exponential and Gaussian describe the tail of the LDF only approximately.

The term $f_\mathrm{int}$, which includes constants from the interaction models, has weaker dependence on the shower geometry and scales with $h_\mathrm{max}$ due to the non-isotropical atmosphere.
Simplifying ${f_\mathrm{int} = 1}$ one still conserves the correlation between the slope of $\mathcal{E}$ and $h_\mathrm{max}$ (see Fig.~\ref{fig:cr-xmax-geo}),
what means, that the radio signal has a high sensitivity to the position of the shower maximum.
However, this one-dimensional slope method does not give additional information on the type of primary particle, which would go beyond $X_\mathrm{max}$.


One can see from Eq.~(\ref{eq1}), that the height of the shower maximum $h_\mathrm{max}$ is encoded in the slope of the LDF. 
what means, that the radio signal has a high sensitivity to the position of the shower maximum.
On the other side, this one-dimensional slope method does not give additional information on the type of primary particle.

\subsection{Systematic uncertainties}
\label{sec:syst_unc}
To study uncertainties given by hadronic interactions and shower-to-shower fluctuations, we performed simulations with recently released CORSIKA~v7.5.
Both, QGSJET-II.04 and EPOS-LHC yield almost the same radio amplitude with a difference less than one given by shower-to-shower fluctuations.

For the detailed study we use QGSJET-II.04, with which we simulated events with parameters similar to the events reconstructed in 2012-2014 with four different primary particles: proton, helium, nitrogen and iron.
The energy resolution (taking noise into account) for each particle is about 10\%, while the shift in the absolute values of the energy is about 12\% between proton and iron.
This is due to the fraction of the primary energy going into the electromagnetic cascade, the same feature is also observed with the optical methods, such as air-Cherenkov and fluorescence.
This shift is much larger, than shower-to shower fluctuations for this particles: 5\% and 1.3\% for proton and iron, respectively.
The reconstruction of the absolute value of the shower maximum is not much affected by the primary composition, since it is reconstructed with the simple slope method, and the one-dimensional slopes are the same for showers of different primary particles with the same shower maxima.

\subsection{Hints from the charge-excess asymmetry}
\begin{figure}[t]
\includegraphics[width=1.0\linewidth]{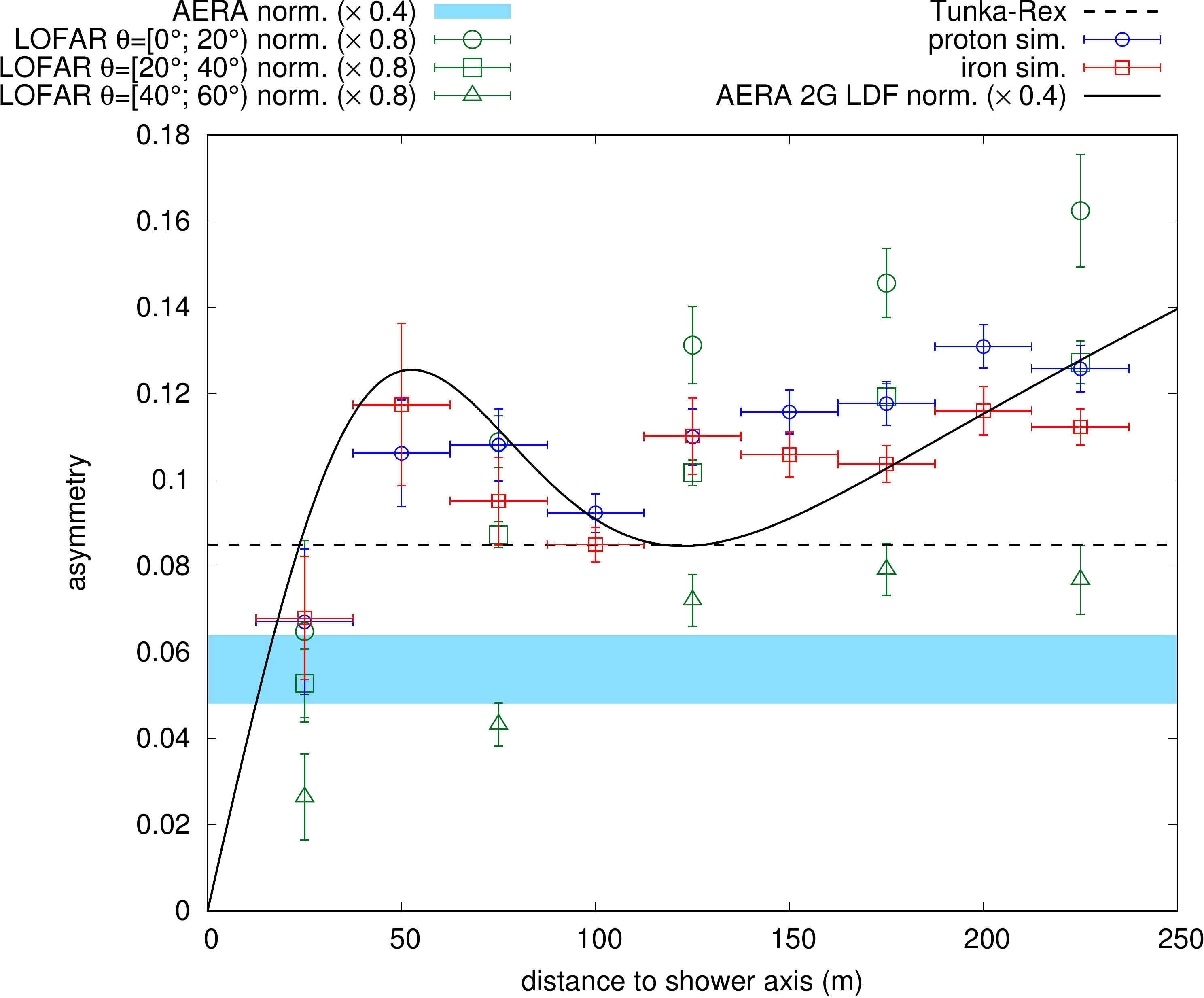}
\caption{
Askaryan asymmetry normalized to the geomagnetic field in the Tunka Valley as a function of the distance to the shower axis. 
Points indicate the polarization measurements by LOFAR~\cite{Schellart:2014oaa} (green) and CoREAS simulations~\cite{Kostunin:2015taa} (blue and red).
The black solid line indicates the LDF asymmetry $\varepsilon(r)$ from Eq.~(\ref{eq_eps}).
The blue band is the polarization measurements by AERA~\cite{Aab:2014esa} (with uncertainties), the dashed line is the theoretical prediction for Tunka-Rex~\cite{Kostunin:2015taa}.}
\label{fig:asymm_profile}
\end{figure}

In the work~\cite{Kostunin:2015taa} it is shown that the charge-excess asymmetry has a non-trivial dependence on the distance to the shower axis, 
particularly, the asymmetry features a local maximum depending on the distance to the shower maximum (see Fig.~\ref{fig:asymm_profile}).
This structure was obtained by analyzing the polarization of CoREAS simulations at individual antenna positions.

As it is mentioned above, in the Tunka-Rex reconstruction the LDF is corrected for charge-excess asymmetry~$\varepsilon$.
To show the consistency of the polarization and LDF approaches, we express $\varepsilon$ in terms of the LDF.
Since the Tunka-Rex LDF is azimuthal symmetric, we have used LDF developed for AERA~\cite{Aab:2015vta} experiment.
The expression for asymmetry has the following form:
\begin{equation}
\varepsilon(r) = \sin\alpha_\mathrm{g}\frac{{\mathcal{E}_{\mathrm{2G}}(r)} - \tilde{\mathcal{E}}_{\mathrm{2G}}(r)}{\mathcal{E}_{\mathrm{2G}}(r) + \tilde{\mathcal{E}}_{\mathrm{2G}}(r)}\,,\,\,\,
\label{eq_eps}
\end{equation}
with
\begin{equation}
\begin{cases}
\mathcal{E}_{\mathrm{2G}}(r) = \mathcal{E}_{\mathrm{2G}}(\bm{r}_x) \\
\tilde{\mathcal{E}}_{\mathrm{2G}}(r) = \mathcal{E}_{\mathrm{2G}}(-\bm{r}_x) \\
\end{cases}\,,
\end{equation}
where $\mathcal{E}_{\mathrm{2G}}(r)$ is the AERA parameterization and vector $\bm{r}_x$ is along the Lorentz force (perpendicular to the shower axis and geomagnetic field).
Taking mean values for the parameters of $\mathcal{E}_{\mathrm{2G}}(r)$, one can obtain the corresponding curve $\varepsilon(r)$ and compare it to one obtained with the polarization approach.
The comparison is presented in Fig.~\ref{fig:asymm_profile}.
One can see, that both definitions are in good agreement, which leads to an interesting conclusion: the asymmetry (or charge-excess) information can be extracted from the more simple measurement of the total radio amplitude, instead of precise measurements of the components of the electrical field.
Measuring the amplitude asymmetry requires higher number of stations per events, but lower signal-to-noise ratios.

As it was shown in Ref.~\cite{Kostunin:2015taa}, the behavior of the asymmetry is connected to the distance to the shower maximum,
i.e. an accurate measurement of the asymmetry by either means should be sensitive to the mass composition.
The idea of an one-antenna analysis can be also applied to a polarization study of the asymmetry: knowing the geometry of the air-shower and the behavior of $\varepsilon(r)$ one can study the mean shower maximum via the mean asymmetry.

Finally, the asymmetry contains information not only on the total number of the charge particles, but also on the dynamics of their creation.

\subsection{Signal recognition with neural networks}
The recognition of the signal and the determination of its amplitude are the most fundamental problems of the basic event reconstruction.
Since the measured signal is a sum of a true radio signal from the air-shower and the background of environment and hardware, the quality of the signal reconstruction is a function of the SNR.
When the SNR is relatively small (conditionally, $\mathrm{SNR}<100$), the influence of the background cannot be neglected.
On average, the amplitude of the measured signal is higher than the amplitude of the true one, and the average fraction between them is expressed as ${\mathcal{E}_\mathrm{t} = \mathcal{E}_\mathrm{m}\sqrt{1 - k/\mathrm{SNR}}}$, where $k$ is a constant, which depends on the definition of SNR.
However, this amplitude can be lower, since the background is uncorrelated with the signal~\cite{Allan:1970xr,Schroeder2012S238}.

In this year we started the investigation of the applicability of neural networks for the signal reconstruction.
We designed a neural network, which gets input traces of 200 counts and predicts the amplitude of the true signal.
We prepared a dataset of about 10000 events, randomly divided in two parts, and used the first one for the training and the second one for the control check.

The control check has shown that the resolution of amplitude reconstruction is about 22\%, which corresponds our standard reconstruction for signals near threshold.
Thus, further investigations are required before this neutral-network approach can be implemented in our standard reconstruction.

\section{Conclusion}
The Tunka Radio Extension is a modern experiment which measures radio emission from air-showers induced by primary cosmic rays with energies above 100~PeV.
Tunka-Rex has proven the feasibility and competitiveness of the radio detection technique.
Operated as a sparse array with spacing of about 200~m between antennas, it has reached a precision of 15\% and 40~g/cm\textsuperscript{2} for the primary energy and the depth of the shower maximum, respectively.

After being upgraded during the last two yeas, Tunka-Rex has now reached triple of its original density and a new trigger from the recently deployed scintillator array Tunka-Grande has been implemented.
This increases the operation time and quality of events, e.g. we now expect more than 1000 events per year instead of about 100 during the first stages of operation.

\section*{Acknowledgements}
The construction of Tunka-Rex was funded by the German Helmholtz association and the Russian Foundation for Basic Research (grant HRJRG-303). 
Moreover, this work has been supported by the Helmholtz Alliance for Astroparticle Physics (HAP), by Deutsche Forschungsgemeinschaft (DFG) grant SCHR 1480/1-1, and by the Russian grant RSF 15-12-20022. 

\bibliography{references}

\end{document}